\documentclass[twoside,leqno,twocolumn]{article}
\usepackage{tikz}
\usepackage{verbatim}
\usepackage{ltexpprt}
\usepackage{amsmath,amssymb} % for \begin{bmatrix}, \begin{smallmatrix}, \begin{align} etc.
\usepackage{amsfonts} % for \mathbb{} etc. `blackboard' for uppercase only.
\usepackage{exscale}
\usepackage{graphicx}
\usepackage{epsfig}
\usepackage{amssymb} % for \begin{bmatrix}, \begin{smallmatrix}, \begin{align} etc.
\usepackage{amsfonts} % for \mathbb{} etc. `blackboard' for uppercase only.
\usepackage[]{hyperref}
\hypersetup{%
linkcolor=blue,anchorcolor=blue,%
citecolor=blue,filecolor=blue,%
menucolor=blue,%
urlcolor=blue,%
colorlinks=true}

\begin{document}

\title{\Large On spectral partitioning of signed graphs\thanks{A preliminary version posted at arXiv.}
}
\author{Andrew Knyazev%
\thanks{%Organization\\ Address\\ Address\\ Email}%
Mitsubishi Electric Research Laboratories (MERL).
%  201 Broadway, 8th floor\\
  Cambridge, MA 02139-1955. 
  Email: knyazev@merl.com}%
}
\date{}
\maketitle

% Copyright Statement
% When submitting your final paper to a SIAM proceedings, it is requested that you include 
% the appropriate copyright in the footer of the paper.  The copyright added should be 
% consistent with the copyright selected on the copyright form submitted with the paper.
% Please note that "20XX" should be changed to the year of the meeting.

% Default Copyright Statement
\fancyfoot[R]{\footnotesize{\textbf{Copyright \textcopyright\ 2018 by Mitsubishi Electric Research Laboratories (MERL)\\
Unauthorized reproduction of this article is prohibited}}}

% Depending on which copyright you agree to when you sign the copyright form, the copyright 
% can be changed to one of the following after commenting out the default copyright statement
% above.

%\fancyfoot[R]{\footnotesize{\textbf{Copyright \textcopyright\ 20XX\\
%Copyright for this paper is retained by authors}}}

%\fancyfoot[R]{\footnotesize{\textbf{Copyright \textcopyright\ 20XX\\
%Copyright retained by principal author's organization}}}

\begin{abstract} \small\baselineskip=9pt
We argue that the standard graph Laplacian is preferable for spectral partitioning of signed graphs compared to the signed Laplacian. Simple examples demonstrate that partitioning based on signs of components of the leading eigenvectors of the signed Laplacian may be meaningless, in contrast to partitioning based on the Fiedler vector of the standard graph Laplacian for signed graphs.  We observe that negative eigenvalues are beneficial for spectral partitioning of signed graphs, making the Fiedler vector easier to compute.  
\end{abstract}
%\begin{keywords}
%spectral clustering, signed graph, signed Laplacian, mass-spring vibration
%\end{keywords}
%\begin{AM}
%05C50, %Combinatorics: Graphs and linear algebra (matrices, eigenvalues, etc.)
%05C70, %Combinatorics: Graphs: Factorization, matching, partitioning, covering and packing
%%05C81, %Combinatorics: Graphs: Random walks on graphs
%15A18, %Linear and multilinear algebra; matrix theory: Eigenvalues, singular values, and eigenvectors
%%15B57, %Linear and multilinear algebra; matrix theory: Hermitian, skew-Hermitian, and related matrices
%58C40, %Global analysis, analysis on manifolds: Spectral theory; eigenvalue problems
%65F15, %Numerical analysis: Numerical linear algebra: Eigenvalues, eigenvectors
%65N25, %Numerical analysis: Partial differential equations, boundary value problems: Eigenvalue problems
%62H30, %Statistics: Classification and discrimination; cluster analysis
%%68T10, %Computer science Pattern recognition, speech recognition 
%91C20. %Game theory, economics, social and behavioral sciences: Clustering
%\end{AM}

%\begin{DOI}
% Place for Digital Object Identifier   
%\end{DOI}

\section{Background and Motivation}\label{s:bm}%
Spectral clustering groups together related data points and separates unrelated data points, using spectral properties of matrices associated with the weighted graph, such as graph adjacency and Laplacian matrices; see, e.g.,\ \cite{chapter_sc,Luxburg2007,Meila01learningsegmentation,ng2002spectral,Shi:2000:NCI:351581.351611,doi:10.1137/0611030,Bolla2013,7023445}. 
The graph Laplacian matrix is obtained from the graph adjacency matrix that represents graph edge weights describing similarities of graph vertices.
The graph weights are commonly defined using a function measuring distances between data points, where the graph vertices represent the data points and the graph edges are drawn between pairs of vertices, e.g.,\ if the distance between the corresponding data points has been measured. %The graph edges are associated with weights, e.g.,\ inversely proportional to the distances. 

Classical spectral clustering bisections the graph according to the signs of the components of the Fiedler vector defined as the eigenvector of the graph Laplacian, constrained to be orthogonal to the vector of ones, and corresponding to the smallest eigenvalue; see \cite{fiedler1973algebraic}.%,fiedler1975property}.

Some important applications, e.g.,\ Slashdot Zoo \cite{Kunegis:2009:SZM:1526709.1526809} and correlation \cite{bansal2004} clustering, naturally lead to signed graphs, i.e., with both positive and \emph{negative} weights. Negative values in the graph adjacency matrix result in more difficult spectral graph theory; see, e.g.,\ \cite{doi:10.1137/130913973}. 

Applying the original definition of the graph Laplacian to signed graphs breaks many useful properties of the graph Laplacian, e.g.,\ leading to negative eigenvalues, making the definition of the Fiedler vector ambivalent. 
The row-sums of the adjacency matrix may vanish, invalidating the definition of the normalized Laplacian. 
These difficulties can be avoided in the \emph{signed Laplacian}, e.g., \cite{2016arXiv160104692G,Kolluri:2004:SSR:1057432.1057434,doi:10.1137/1.9781611972801.49}, defined similarly to the graph Laplacian, but with the diagonal entries positive and large enough to make the signed Laplacian positive semi-definite.%, allowing one to formally reuse the definitions of the Fiedler vector and the normalized Laplacian. 

We argue that the original graph Laplacian is a more natural and beneficial choice, compared to the popular signed Laplacian, for spectral partitioning of signed graphs. We explain why the definition of the Fiedler vector should be based on the smallest eigenvalue, no matter whether it is positive or negative, motivated by the classical model of transversal vibrations of a mass-spring system, e.g., \cite{gould,Demmel99}, but with some springs having negative stiffness, cf. \cite{AKnegativePatent}.

Inclusions with negative stiffness can occur in mechanics if the inclusion is stored with energy \cite{natureNegativeStiffness2001}, e.g.,\ pre-stressed and constrained. We~design inclusions with negative stiffness by pre-tensing the spring to be repulsive \cite{CHRONOPOULOS201748}. Allowing only the transversal movement of the masses, as in \cite{Demmel99}, gives the necessary constraints. 

The~resulting eigenvalue problem for the vibrations remains mathematically the same, for the original graph Laplacian, no matter if some entries in the adjacency matrix of the graph are negative. In~contrast, to motivate the signed Laplacian, the ``inverting amplifier'' model in \cite[Sec. 7]{doi:10.1137/1.9781611972801.49}  uses a questionable argument, where the sign of negative edges changes in the denominator of the potential, but not in its numerator  

Turning to justification of spectral clustering via relaxation, we compare the 
standard ``ratio cut,'' e.g.,\ \cite{Meila01learningsegmentation,ng2002spectral},
and ``signed ratio cut'' of \cite{doi:10.1137/1.9781611972801.49}, noting that minimizing the signed ratio cut 
may amplify cutting positive edges. 
We illustrate the behavior of the Fiedler vector for an intuitively trivial case of partitioning a linear graph modelled by vibrations of a string. We~demonstrate numerically and analyze deficiencies of the signed Laplacian vs. the standard Laplacian for spectral clustering on a few simple examples.

Graph-based signal processing introduces eigenvectors of the graph Laplacian as natural substitutions for the Fourier basis. The construction of the graph Laplacian of~\cite{knyazev2015conjugate} is extended in \cite{knyazev2015edge} to the case of some negative weights, leading to edge enhancing denoising of an image that can be used as a precursor for image segmentation along the edges. We extend the use of negative weights to graph partitioning in the present paper. 

 The rest of the paper is organized as follows. We introduce spectral clustering in Section \ref{s:isc} via eigendecomposition of the graph Laplacian. Section \ref{s:string} deals with a simple, but representative, example---a linear graph,---and motivates spectral clustering by utilizing properties of low frequency mechanical vibration eigenmodes of a discrete string, as an example of a mass-spring model. Negative edge weights are then naturally introduced in Section \ref{s:n} as corresponding to repulsive springs, and the effects of negative weights on the eigenvectors of the Laplacian are informally predicted by the repulsion of the masses connected by the repulsive spring. In Section \ref{s:sl}, we present simple motivating examples, discuss how the original and signed Laplacians are introduced via relaxation of combinatorial optimization, and numerically compare their eigenvectors and gaps in the spectra.
Possible future research directions are spotlighted in Section \ref{s:future}.  

\section{Brief introduction to spectral clustering}\label{s:isc}
Let entries of the real symmetric  $N$-by-$N$ \emph{data similarity} matrix $W$ be called \emph{weighs} and the matrix $D$ be diagonal, made of row-sums of the 
matrix $W$. 
The matrix $W$ may be viewed as a matrix of scores that digitize similarities of pairs  of $N$  data points. Similarity matrices are commonly determined from their counterparts, distance matrices, which consist of pairwise distances between the data points. The similarity is small if the distance is large, and vice versa.  Traditionally, all the weighs/entries in $W$ are assumed to be non-negative, which is automatically satisfied for distance-based similarities. 
We are interested in clustering in a more general case of both positive and negative weighs, e.g.,\ associated with pairwise correlations of the data vectors. 

Data clustering is commonly formulated as graph partitioning, defined on data represented in the form of a graph $G = (V,\, E,\, W)$, with $N$ vertices in $V$ and $M$ edges in $E$, where entries of the $N$-by-$N$ \emph{graph adjacency} matrix $W$ are weights of the corresponding  edges. The~graph is called \emph{signed} if some edge weighs are negative. 
A partition of the vertex set $V$ into subsets generates subgraphs of $G$ with desired properties. 
 
A partition in the classical case of non-weighted graphs minimizes the number of edges between separated sub-graphs, while 
maximizes the number of edges within each of the sub-graphs.
The goal of partitioning of signed graphs, e.g., into two vertex subsets $V_1$ and $V_2$, can be to minimize the total weight of the  positive cut edges, while at the same time to maximize the absolute total weight of the negative cut edges.
For uniform partitioning, one also needs to well-balance sizes/volumes of $V_1$ and $V_2$.
Traditional approaches to graph partitioning are combinatorial and naturally fall under the category of NP-hard problems,
solved using heuristics, such as relaxing the combinatorial constraints. %, as we briefly discuss below in Subsection \ref{ss:relax}.

Data clustering via graph spectral partitioning is a state-of-the-art tool, which is known to produce high quality clusters at reasonable costs of numerical solution of an eigenvalue problem 
for a matrix associated with the graph, e.g.,\ $Lx=\lambda x$ for the graph Laplacian matrix $L=D-W$, where the scalar $\lambda$ denotes the eigenvalue corresponding to the eigenvector~$x$. To~simplify our presentation for the signed graphs, we mostly avoid the normalized  Laplacian $D^{-1}L=I-D^{-1}W$, where $I$ is the identity matrix, e.g., since $D$ may be singular.

The Laplacian matrix $L$ always has the number $0$ as an eigenvalue; and 
the column-vector of ones is always a trivial eigenvector of $L$ corresponding to the zero eigenvalue.  
Since the graph adjacency matrix $W$ is symmetric, the graph Laplacian matrix is also symmetric, so all eigenvalues of $L$ are real and the various eigenvectors can be chosen to be mutually orthogonal. 
All eigenvalues are non-negative if the graph weights are all non-negative.

A nontrivial eigenvector of the matrix $L$ corresponding to smallest eigenvalue $\lambda$ of $L$, commonly called the Fiedler vector after the author of \cite{fiedler1973algebraic}, bisects the graph $G$ into only two parts, according to the signs of the entries of the eigenvector. Since the Fiedler vector, as any other nontrivial eigenvector, is orthogonal to the vector of ones, it must have entries of opposite signs, thus, the sign-based bisection always generates a non-trivial two-way graph partitioning. We explain in Section \ref{s:string}, why such a partitioning method is intuitively meaningful.

A multiway spectral partitioning is obtained from ``low frequency eigenmodes,'' i.e., eigenvectors corresponding to a cluster of smallest eigenvalues, of the  Laplacian matrix $L.$ The cluster of (nearly)-multiple eigenvalues naturally leads to the need of considering a Fiedler invariant subspace of $L$, spanned by the corresponding eigenvectors, extending the Fiedler vector, since the latter may be not unique or well defined numerically in this case. The Fiedler invariant subspace provides a geometric embedding of graph's vertices, reducing the graph partitioning problem to the problem of clustering of a point cloud of embedded graph vertices in a low-dimensional Euclidean space. However, the simple sign-based partitioning from the Fiedler vector has no evident extension to the Fiedler invariant subspace. 

Practical multiway spectral partitioning can be performed using various competing heuristic algorithms, greatly affecting the results. 
While these same heuristic algorithms can as well be used in our context of signed graphs, for clarity of presentation we restrict ourselves in this work only to two-way partitioning using the component signs of the Fiedler vector. 

The presence of negative weights in signed graphs brings new challenges to spectral graph partitioning: 
\begin{itemize}
 \item negative eigenvalues of the graph Laplacian make the definition of the Fiedler vector ambiguous, e.g., whether the smallest negative or positive eigenvalues, or may be the smallest by absolute value eigenvalue, should be used in the definition; 
 \item difficult spectral graph theory, cf. \cite{2016arXiv160104692G} and \cite{Luxburg2007};
 \item possible zero diagonal entries of the degree matrix $D$ in the normalized  Laplacian $D^{-1}L$, cf. \cite{Shi:2000:NCI:351581.351611};
 \item violating the maximum principle---the cornerstone of a theory of connectivity of clusters \cite{fiedler1973algebraic}; 
 \item breaking the connection of spectral clustering to random walks and Markov chains, cf. \cite{Meila01learningsegmentation};
 \item the quadratic form $x^\mathsf{T}Lx$ is not ``energy,'' e.g., in the heat (diffusion) equation; cf. a forward-and-backward diffusion in \cite{1021076,Tang2016};
 \item the graph Laplacian can no longer be viewed as a discrete analog of the Laplace-Beltrami operator on a Riemannian manifold that motivates spectral manifold learning; e.g.,\  \cite{Ham:2004:KVD:1015330.1015417,Rossi2015}.
\end{itemize}

Some of these challenges can be addressed by defining a \emph{signed} Laplacian as follows. 
Let the matrix $\bar{D}$ be diagonal, made of row-sums of the \emph{absolute values of the entries} of the matrix $W$, which thus are positive, so that $\bar{D}^{-1}$ is well-defined. We define the \emph{signed} Laplacian $\bar{L}=\bar{D}-W$ 
following, e.g., \cite{2016arXiv160104692G,Kolluri:2004:SSR:1057432.1057434,doi:10.1137/1.9781611972801.49}. The signed Laplacian is positive semi-definite, with all eigenvalues non-negative. The Fiedler vector of the signed Laplacian is defined in \cite{2016arXiv160104692G,Kolluri:2004:SSR:1057432.1057434,doi:10.1137/1.9781611972801.49} as an eigenvector corresponding to the smallest eigenvalue and different from the trivial constant vector. We finally note recent work \cite{doi:10.1137/16M1082433}, although it is not a part of our current investigation. 

In the rest of the paper, we justify spectral partitioning of signed graphs using the original definition of the graph Laplacian $L$, and argue that better quality clusters can generally be expected from eigenvectors of the original $L$, rather than from the signed  Laplacian $\bar{L}$.  We use the intuitive mass-spring model to explain novel effects of negative stiffness or spring repulsion on eigenmodes of the standard Laplacian, but we are unaware of a physical model for the signed Laplacian. 

\begin{figure}
\centering
\hspace{-20mm}
\includegraphics[width=0.55\linewidth,height=0.26\linewidth]{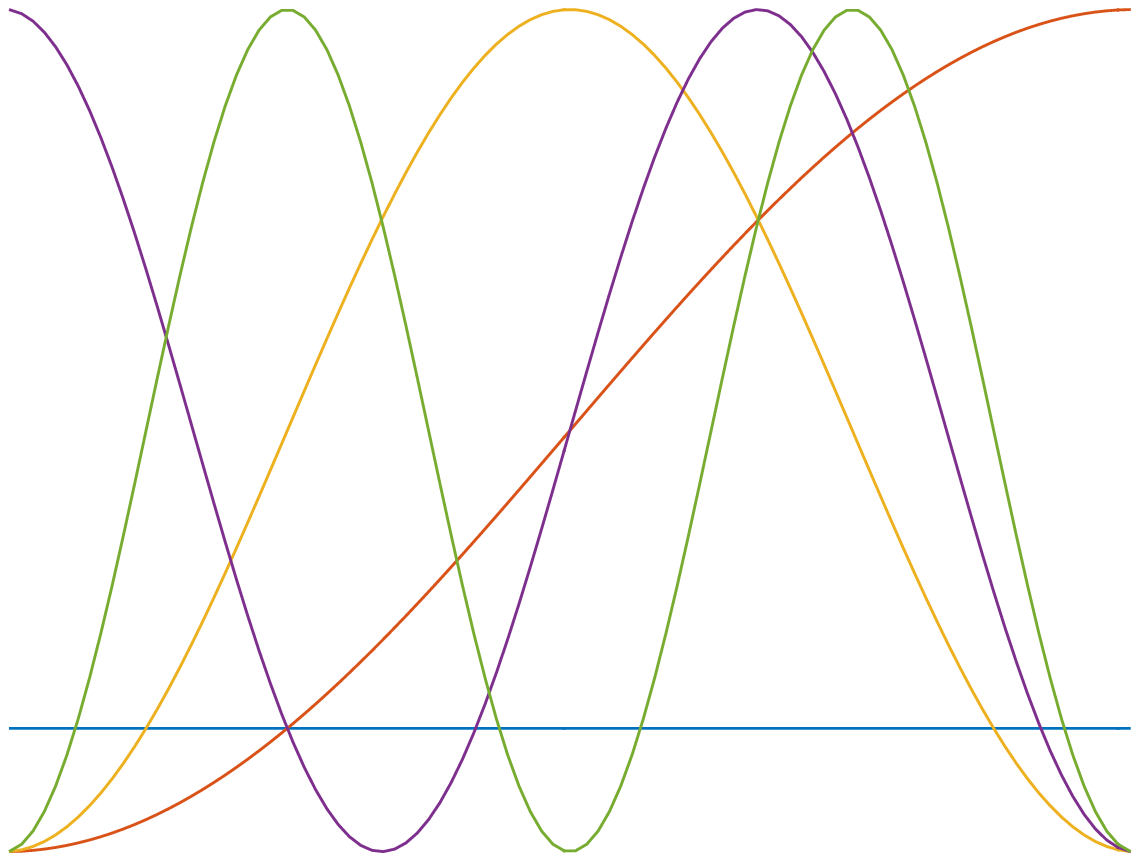} 
\hspace{-5mm}
\includegraphics[width=0.55\linewidth,height=0.26\linewidth]{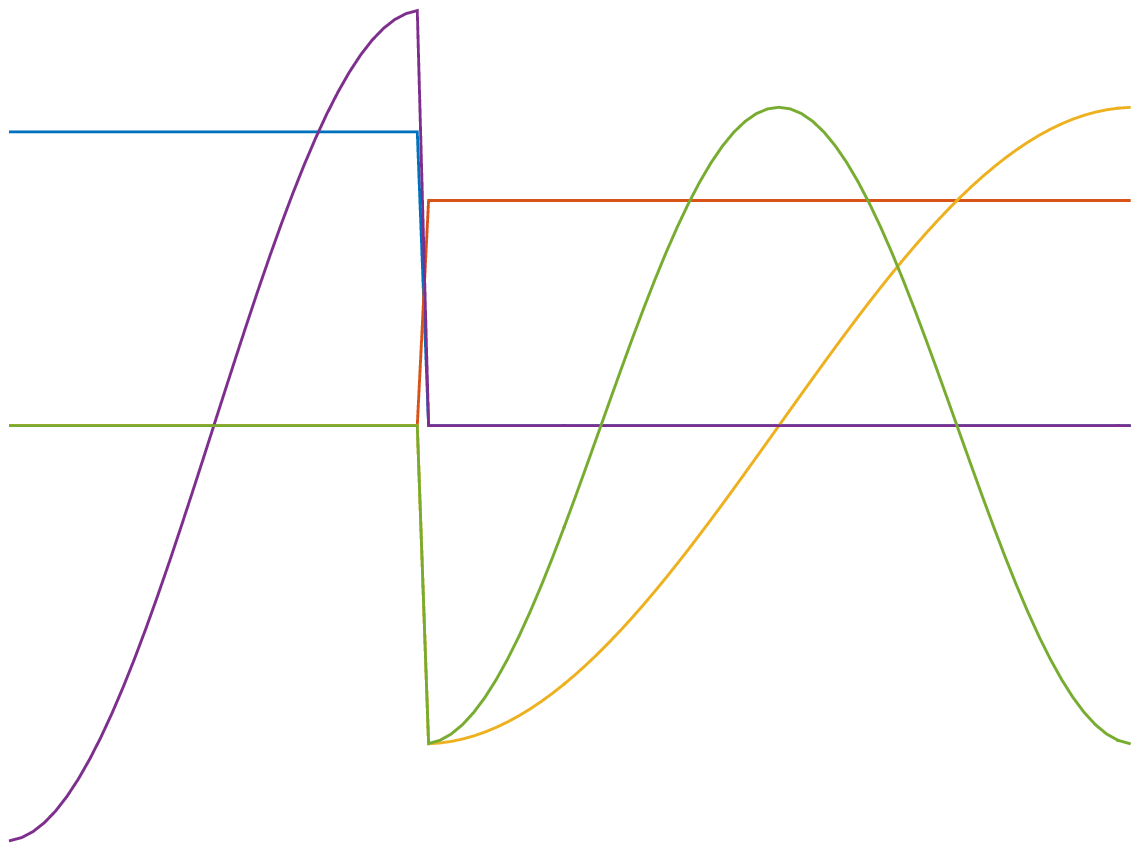}
\hspace{-20mm}
\caption{Low frequency eigenmodes of a string (left) and two disconnected pieces of the string (right).%
}
\label{fig:2}
\end{figure}

\section{Linear graph Laplacian and low frequency eigenmodes of a string}\label{s:string}
Spectral clustering can be justified intuitively via a well-known identification of the graph Laplacian matrix $L$ with a classical problem of vibrations of a mass-spring system without boundary conditions, with $N$ masses and $M$ springs, where the stiffness of the springs is related to the weights of the graph; see, e.g.,~\cite{Park20143245}. References \cite{PASTERNAK20146676,Park20143245} consider lateral vibrations, where \cite{PASTERNAK20146676} allows springs with negative stiffness. We prefer the same model, but with transversal vibrations, as in \cite{Demmel99}, although the linear eigenvalue problem is the same, for the original graph Laplacian, no matter whether the vibrations are lateral or transversal, under the standard assumptions of infinitesimal displacements from the equilibrium and no damping. The transversal model allows relating the linear mass-spring system to the discrete analog of an ideal string  \cite[Fig.~2]{gould} and provides the necessary constraints for us to introduce a specific physical realization of inclusions with the negative stiffness by pre-tensing some springs to be repulsive.  
We start with the simplest example, where the mass-spring system is a discrete string. 

\subsection{All edges with unit weights}
Let  $w_{i-1\, i}%=w_{i\, i}
= w_{i\, i+1}=1 $
with all other zero entries, so that the graph
Laplacian $L=D-W$ is a tridiagonal matrix 
\begin{equation}\label{e1}
L = \left( \begin{array}{ccccc}
1 & -1 & & &\\
-1 & 2 & -1 & &\\
& \ddots & \ddots & \ddots &\\
& &-1 & 2 & -1 \\
& & & -1 & 1 \end{array} \right) \end{equation}
that has nonzero entries 
$1$ and $-1$ in the first row, $-1$ and $1$ in the last row, and 
$[-1\,\; 2\, -1]$ in every other row---a standard finite-difference 
approximation of the negative second derivative of functions with vanishing 
first derivatives at the end points of the interval. Its eigenvectors 
are the basis vectors of the discrete cosine transform; see the first five 
low frequency eigenmodes (the eigenvectors corresponding to the smallest eigenvalues) of  $L$
displayed in the left panel in Figure \ref{fig:2}.  
Let us note that  these eigenmodes 
all turn flat at the end points of the interval. 

The flatness is attributed to the vanishing first derivatives, which manifests itself  
in the fact, e.g.,\ that 
the Laplacian row sums always vanish, including in the first and last rows, corresponding to the ``boundary.'' 

\begin{figure}
\centering
\includegraphics[width=0.78\linewidth]{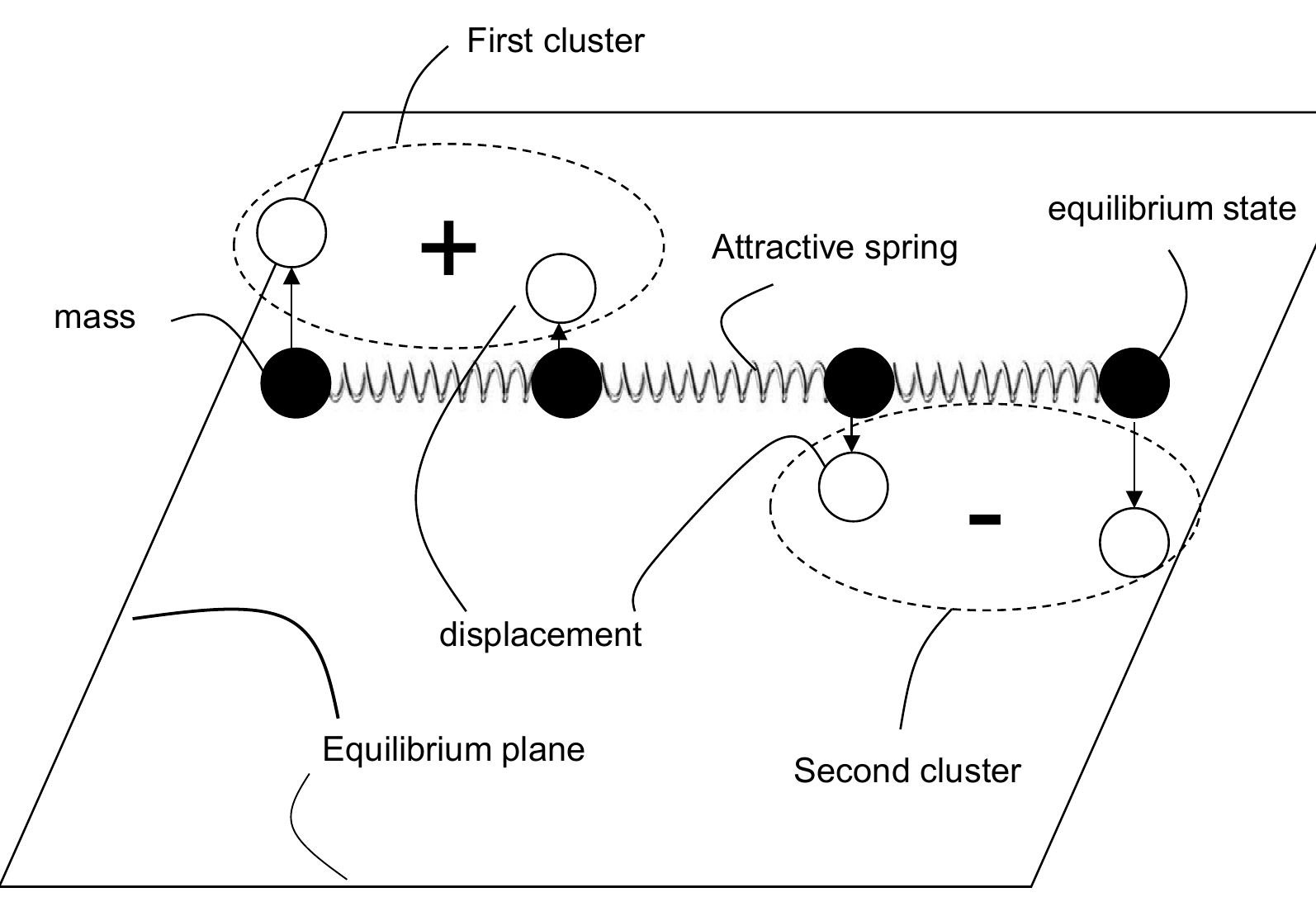}
\caption{Traditional linear mass-spring system.}
\label{fig:m-s-p}
\end{figure}

Eigenvectors of matrix \eqref{e1} are well-known in mechanics, as 
they represent shapes of transversal vibration modes of 
a discrete analog of a string---a linear system of masses connected with springs. 
Figure~\ref{fig:m-s-p} illustrates a system with $N=4$ masses and $M=3$ springs.

The frequencies squared $\omega^2$ of the vibration modes $x$ are the eigenvalues $\lambda\geq0$, e.g.,\  \cite[p. 15]{gould}.
The eigenvectors $x$ of the graph Laplacian can be called \emph{eigenmodes} because of this 
mechanical analogy. The smallest eigenvalues $\lambda=\omega^2$ correspond to low frequencies $\omega$,
explaining the terminology used in the caption in Figure \ref{fig:2}.
Our system of masses is not attached, thus there is always a trivial eigenmode, 
where the whole system goes up/down, i.e., the eigenvector $x$ is constant with the zero frequency/eigenvalue
$\omega^2=\lambda=0$. 

If the system consists of $k$ completely separate components, each component can independently move up/down 
in zero frequency vibration, resulting in total $k+1$ multiplicity of the zero frequency/eigenvalue, where the corresponding eigenvectors are all piecewise constant with discontinuities between the components. Such a system represents a graph consisting of $k$ completely separate sub-graphs and can be used to motivate $k$-way spectral partitioning. 

In our case $k=2$, the Fiedler vector is chosen orthogonal to the trivial constant eigenmode, and thus is not only piecewise constant, but also has strictly positive and negative components, determining the two-way spectral partitioning. 

Figure \ref{fig:m-s-p} shows transversal displacements of the masses from the equilibrium plane for the first nontrivial mode, which is the Fiedler vector,
where  the two masses on the left side of the system move synchronously up, while the two masses on the right side of the system move synchronously down. This is the same eigenmode as drawn in red color in Figure  \ref{fig:2} left panel for a similar linear system with a number of masses large enough to visually appear as a continuous string.  Performing the spectral bisection (two-way partitioning) according to the signs of the Fiedler vector, one puts the data points corresponding to the masses in the left half of the mass-spring system into one cluster and those in the right half into the other. The Fiedler vector is not piecewise constant, since the partitioned components are not completely separate. 

The amplitudes of the Fiedler vector components are also very important. The  amplitude of the component squared after proper scaling can be interpreted as a probability of the corresponding data point to belong to the cluster determined according to the sign of the component.  For example, the Fiedler vector in Figure~\ref{fig:m-s-p} has small absolute values of its components in the middle of the system. With the number of masses increased, the  components in the middle of the system approach zero. Perturbations of the graph weights may lead to the sign changes in the small components, putting the corresponding data points into a different cluster. 

\subsection{A string with a single weak link (small edge weight)}
Next, we set a very small value $w_{i\, i+1}=w_{i+1 \, i}$ for some index~$i$, keeping all other entries of the matrix $W$ the same as before.  In terms of clustering, this example represents a situation where there is an intuitively evident bisection with one cluster containing all data points with indexes $1,\ldots,i$ and the other with $i+1,\ldots,N$. 
In terms of our mass-spring system interpretation, we have a discrete string with one weak link, i.e.,
one spring with such a small stiffness that makes two pieces of the string nearly disconnected. 

Let~us check how the low frequency eigenmodes react to such a change.
The first five vectors of the 
corresponding Laplacian are shown in Figure \ref{fig:2}, right panel.  
We~observe that all the eigenvectors plotted in Figure \ref{fig:2} are aware of softness (small stiffness) of the spring between the masses with the indexes $i$ and $i+1$. Moreover, their behavior around the soft spring is very specific---they are all flat on both sides of the soft spring! 

The presence of the flatness in the low frequency modes of the graph Laplacian $L$ on both sides of the soft spring is easy to explain mathematically. When the value $w_{i\, i+1}=w_{i+1 \, i}$ is small relative to other entries, the matrix  $L$ becomes 
nearly block diagonal, with two blocks that approximate the graph Laplacian matrices on sub-strings  
to the left and right of the soft spring. 
The low frequency eigenmodes
of the graph Laplacian $L$ thus approximate combinations of the 
low frequency eigenmodes of the graph Laplacians on the sub-intervals.

However, each of the low frequency eigenmodes of the graph Laplacian on the sub-interval
is flat on both ends of the sub-interval, as explained above. 
Combined, it results in the flatness in the low frequency modes of the graph Laplacian $L$
on both sides of the soft spring.

The flatness is also easy to explain in terms of mechanical vibrations. The soft spring between the masses with the indexes $i$ and $i+1$  makes the masses nearly disconnected, so the system can be well approximated by two independent disconnected discrete strings with free boundary conditions, on the left and on the right to the soft spring. Thus, the low frequency vibration modes of the system are visually discontinuous at the soft spring, and nearly flat on both sides of the soft spring. 

Can we do better and make the flat ends bend in the opposite directions, making it easier to determine the bisection, e.g.,\ using low-accuracy computations of the eigenvectors? 
In \cite{knyazev2015edge}, where graph-based edge-preserving signal denoising is analyzed, 
we have suggested to enhance the edges of the signal by introducing negative edge weights in the graph, cf. \cite{1021076}. 
In the next section, we put a spring which separates the masses by repulsing them and see how the repulsive spring  
affects the low-frequency vibration modes.

\section{Negative weights for spectral clustering}\label{s:n}
In~our~mechanical vibration model of a spring-mass system, the masses that are tightly connected have a tendency to move synchronically in low-frequency free vibrations. Analyzing the signs of the components corresponding to different masses of the low-frequency vibration modes determines the clusters. 

\begin{figure}
\centering
\includegraphics[width=0.88\linewidth]{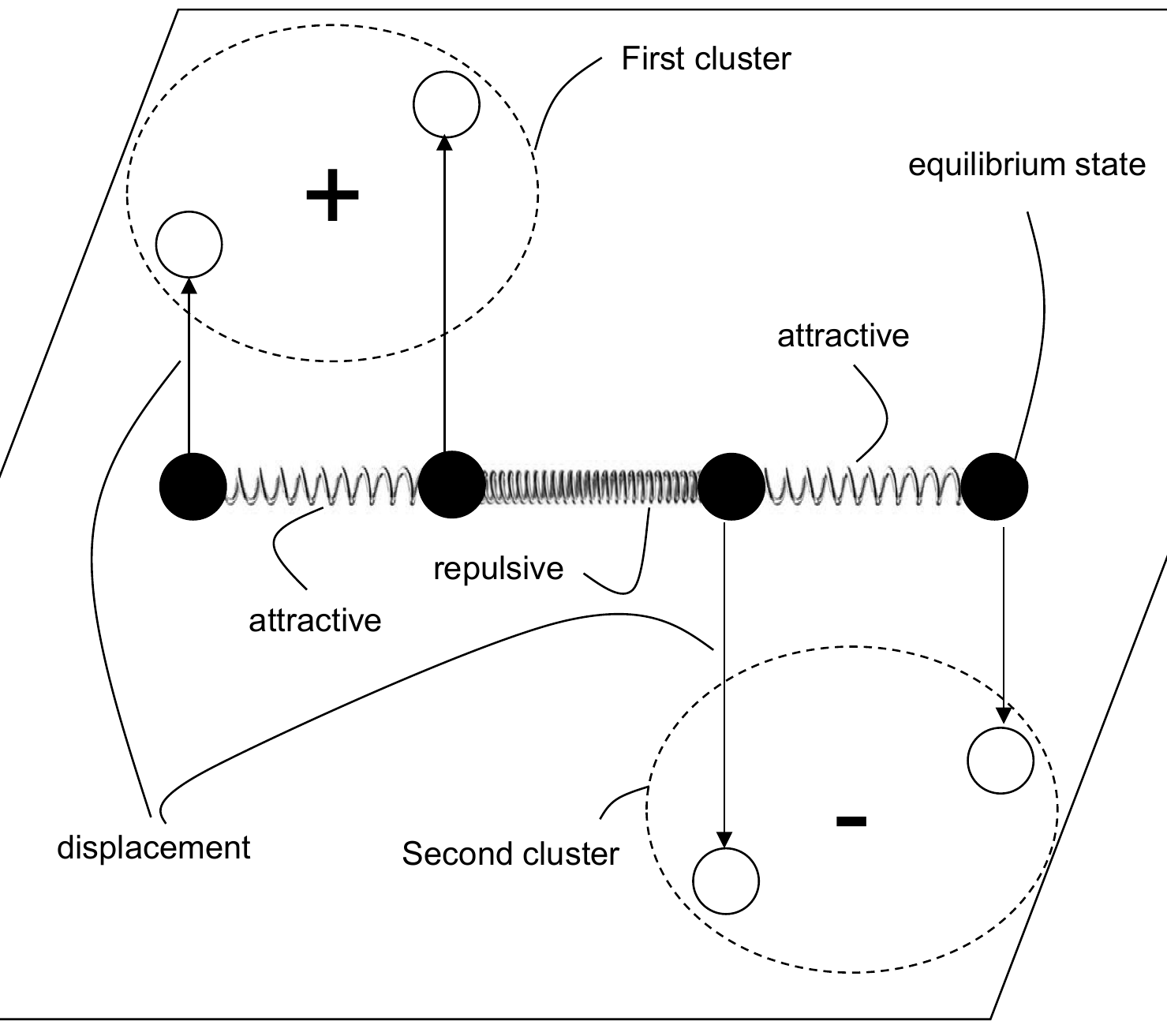}
\caption{Linear mass-spring system with repulsion.}
\label{fig:m-s-n}
\end{figure}

The mechanical vibration model describes conventional clustering when all the springs are pre-tensed to create attracting forces between the masses, where the mass-spring system is subject to transverse vibrations, i.e., the masses are constrained to move only in a transverse direction, perpendicular to a plane of the mass-spring system. 
However, one can also pre-tense some of the springs to create repulsive forces between some pairs of masses, as illustrated in Figure~\ref{fig:m-s-n}. For example, the second mass is connected by the attractive spring to the first mass, but by the repulsive spring to the third mass in Figure~\ref{fig:m-s-n}. The repulsion has no effect in the equilibrium, since the masses are constrained to displacements only in the transversal direction, i.e. perpendicular to the equilibrium plane. 
When the second mass deviates, shown in white circle in Figure~\ref{fig:m-s-n}, from its equilibrium  position, shown in back circle in Figure~\ref{fig:m-s-n}, the transversal component of the attractive force from the attractive spring on the left is oriented toward the equilibrium position, while the transversal component of the repulsive force from the repulsive spring on the right is in the opposite direction, resulting in opposite signs in the equation of the balance of the two forces. Since the stiffness is the ratio of the force and the displacement, the attractive spring on the left has effective positive stiffness, but the repulsive spring represents the inclusion with effective negative stiffness, due to the opposite directions of the corresponding forces.   

In the context of data clustering formulated as graph partitioning, that corresponds to negative entries in the adjacency matrix. The negative entries in the adjacency matrix are not allowed in conventional spectral graph  partitioning. However, the model of mechanical vibrations of the spring-mass system with repulsive springs is still valid, allowing us to extend the conventional approach to the case of negative weights.

The masses which are attracted move together in the same direction in low-frequency free vibrations, while the masses which are repulsed have the tendency to move in the opposite direction. Moreover, the eigenmode vibrations of the spring-mass system relate to the corresponding wave equation, where the repulsive phenomenon makes it possible for the time-dependent solutions of the wave equation to exponentially grow in time, if they correspond to negative eigenvalues. 

Figure \ref{fig:m-s-n} shows the same linear mass-spring system as Figure~\ref{fig:m-s-p}, except that the middle spring is repulsive, bending the shape of the Fiedler vector in the opposite directions on different sides of the repulsive spring. The clusters in Figure~\ref{fig:m-s-p} and Figure \ref{fig:m-s-n} are the same, based on the signs of the Fiedler vectors. However, the data points corresponding to the middle masses being repulsed more clearly belong to different clusters in Figure \ref{fig:m-s-n}, compared to Figure~\ref{fig:m-s-p},
because the corresponding components in the Fiedler vector are larger by absolute value in Figure \ref{fig:m-s-n} vs. Figure~\ref{fig:m-s-p}. Determination of the clusters using the signs of the Fiedler vector 
is easier for larger components, since they are less likely to be computed with a wrong sign due to data noise or inaccuracy of computations, e.g.,\ small number of iterations. 

\begin{figure}
\centering
\hspace{-20mm}
\includegraphics[width=0.55\linewidth,height=0.3\linewidth]{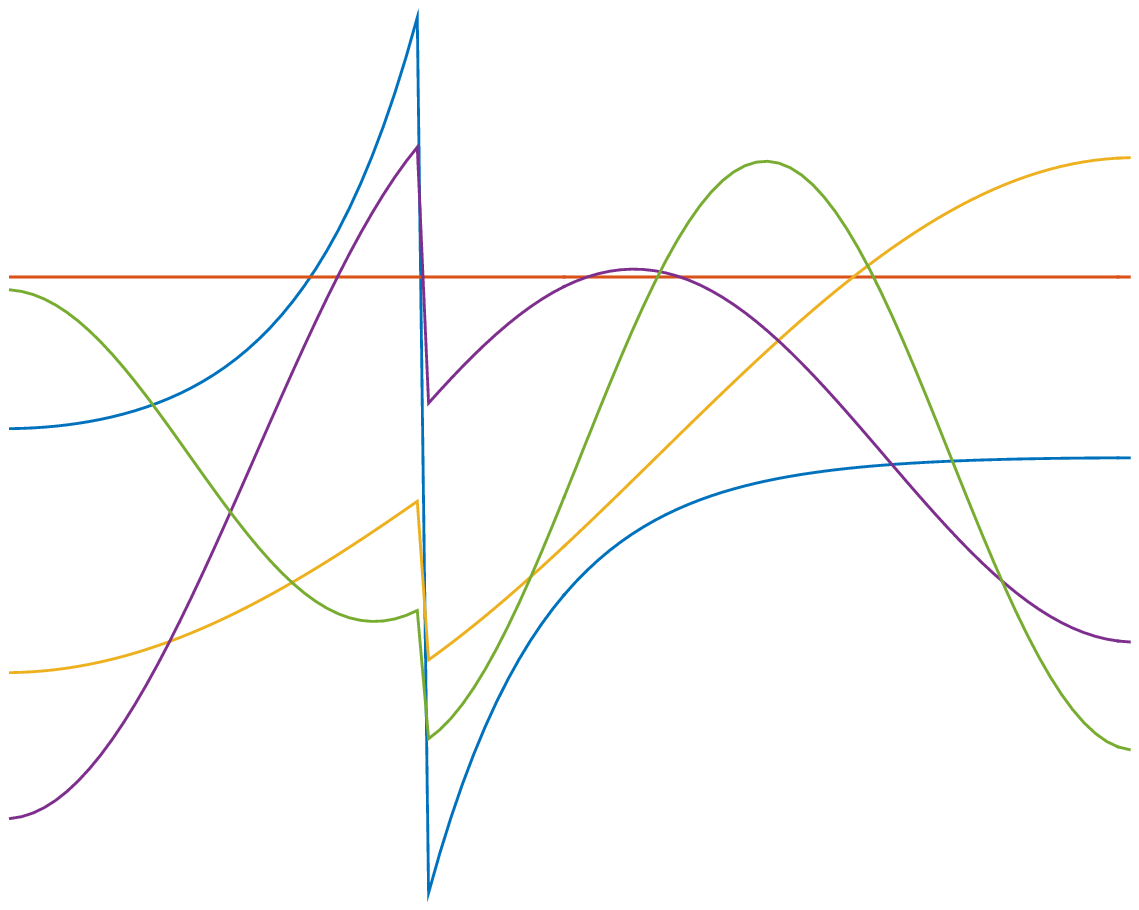}
\hspace{-5mm}
\includegraphics[width=0.55\linewidth,height=0.3\linewidth]{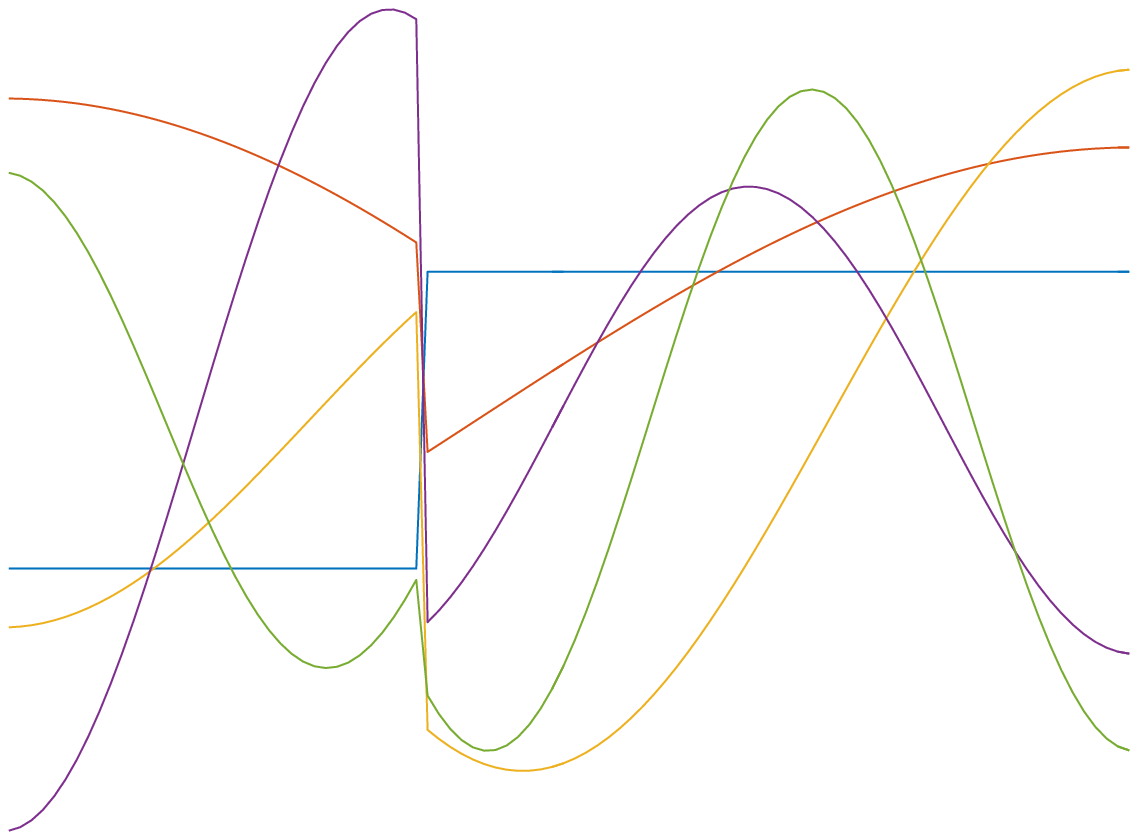}
\hspace{-20mm}
\caption{The same eigenmodes, but negative weights, original (left) and signed (right) Laplacians.}
\label{fig:3}
\end{figure}

Figure \ref{fig:3} left panel displays the five eigenvectors, including the trivial one, for the five smallest eigenvalues of the same tridiagonal graph Laplacian as that corresponding to the right panel in Figure \ref{fig:2}
except that the small positive entry of the weights $w_{i\, i+1}=w_{i+1 \, i}$ for the same $i$  
is substituted by $-0.05$ in Figure \ref{fig:3}. Figure \ref{fig:3} right panel displays the five leading eigenvectors of the corresponding signed Laplacian. The left panel of Figure \ref{fig:3} illustrates the predicted phenomenon of the repulsion, in contrast to the right panel. The Fiedler vector of the Laplacian, displayed in blue color in the left panel of Figure \ref{fig:3}, is most affected by the repulsion compared to higher frequency vibration modes. This effect gets more pronounced if the negative weight increases by absolute value, as we observe in other tests not shown here. 

The Fiedler vector of the signed Laplacian with the negative weight displayed in blue color in the right panel of Figure~\ref{fig:3} looks piecewise constant, just the same as the Fiedler vector of the Laplacian with nearly zero weight shown in red color in Figure \ref{fig:2} right panel. We now prove that this is not a coincidence. 
Let us consider a linear graph corresponding to Laplacian \eqref{e1}. We first remove one of the middle edges and 
define the corresponding graph Laplacian $L_0$. Second, we put this edge back but with the negative unit weight $-1$  and 
define the corresponding signed Laplacian $\bar{L}$. It is easy to verify 
  \begin{equation}\label{e6}
\bar{L}-L_0 = \left( \begin{array}{cccccc}
\cdots & \cdots & \cdots & \cdots\\
 \cdots & 1 & 1 & \cdots\\
 \cdots & 1 & 1 & \cdots\\
\cdots & \cdots & \cdots& \cdots\end{array} \right), \end{equation}
where all dotted entries are zeros.

The Fiedler vector of $L_0$ is evidently piece-wise constant with one discontinuity at the missing edge, since
the graph Laplacian $L_0$ corresponds to the two disconnected discrete string pieces. 
Let $x_0$ denote the  Fiedler vector of $L_0$ shifted by the vector of ones and scaled so that its components with the opposite sign are simply $+1$ and $-1$, while still $L_0x_0=0$.  
We get $(\bar{L}-L_0)x_0=0$ from \eqref{e6}, thus, also $\bar{L}x_0=0$, i.e., $x_0$ is the Fiedler vector of both matrices $\bar{L}$ and $L_0$, where in the latter our only negative weight is simply nullified. 

 \begin{figure}
\centering
\hspace{-9mm}
\includegraphics[width=0.54\linewidth]{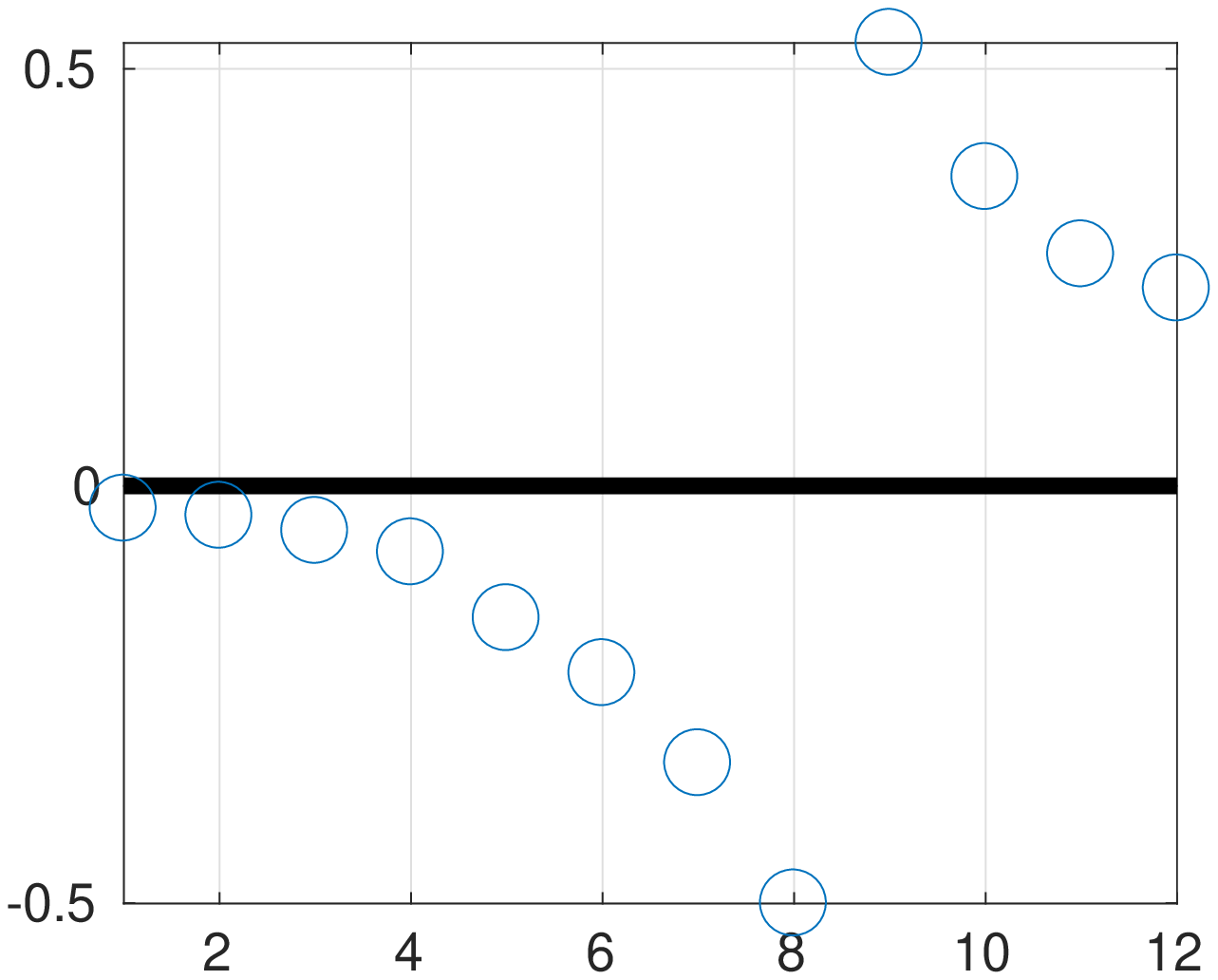}
\hspace{-5mm}
\includegraphics[width=0.54\linewidth]{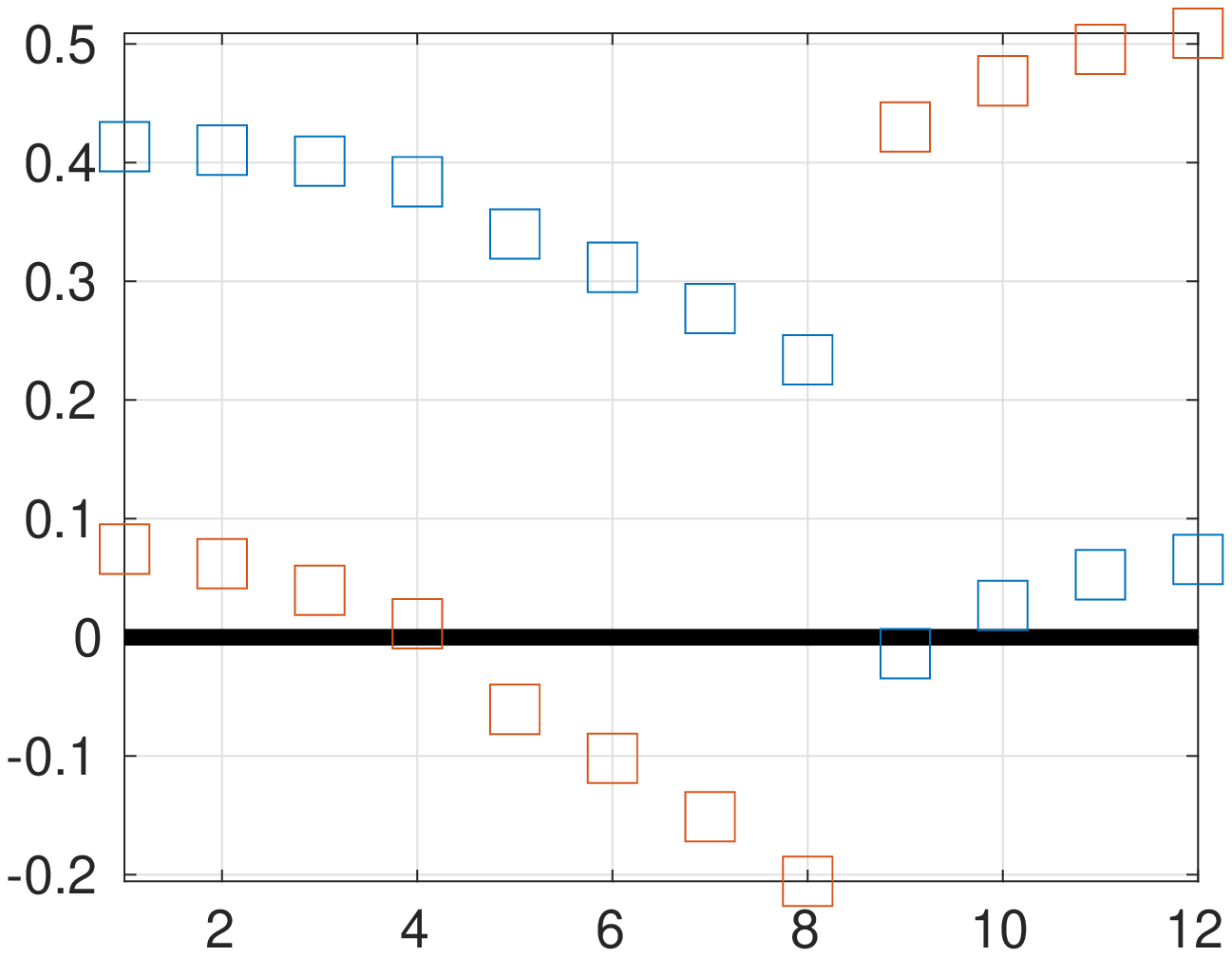}
\hspace{-9mm}
\caption{Laplacian eigenmodes, original (left) and signed (right), a ``noisy'' $12$-mass string with a negative weight at one edge between vertices $8$ and $9$.}
\label{fig:na}
\end{figure}

\section{Comparing the original vs. signed Laplacians}\label{s:sl}
We present a few simple motivating examples, discuss how the original and signed Laplacians are introduced via relaxation of combinatorial optimization, and compare their eigenvectors and gaps in the spectra, computed numerically for these examples.
\subsection{Linear graph with noise}
We~consider another standard linear mass-spring system with $12$ masses and one repulsive spring, $w_{89}=w_{98}=-1/2$ between masses $8$ and $9$, but add to the graph adjacency an extra full random matrix with entries uniformly distributed between $0$ and $10^{-2}$, modelling noise in the data. It turns out that in this example the two smallest eigenvalues of the signed Laplacian form a cluster, making individual eigenvectors unstable with respect to the additive noise, leading to meaningless spectral clustering, if based on the signs on the components of any of the two eigenvectors. Specifically, the exact Laplacian eigenmodes are shown in Figure \ref{fig:na}: the original Fiedler (left panel) and both eigenvectors of the signed Laplacian (right panel). The Fiedler vector of the original Laplacian clearly suggests the perfect cut. Neither the first nor the second (giving it a benefit of a doubt) exact eigenvectors of the signed Laplacian result in meaningful clusters, using the signs of the eigenvector components as suggested  in \cite{doi:10.1137/1.9781611972801.49}.  

\subsection{``Cobra'' graph}
\begin{figure}
%\centering
\hspace{-10mm}
\includegraphics[width=1.15\linewidth]{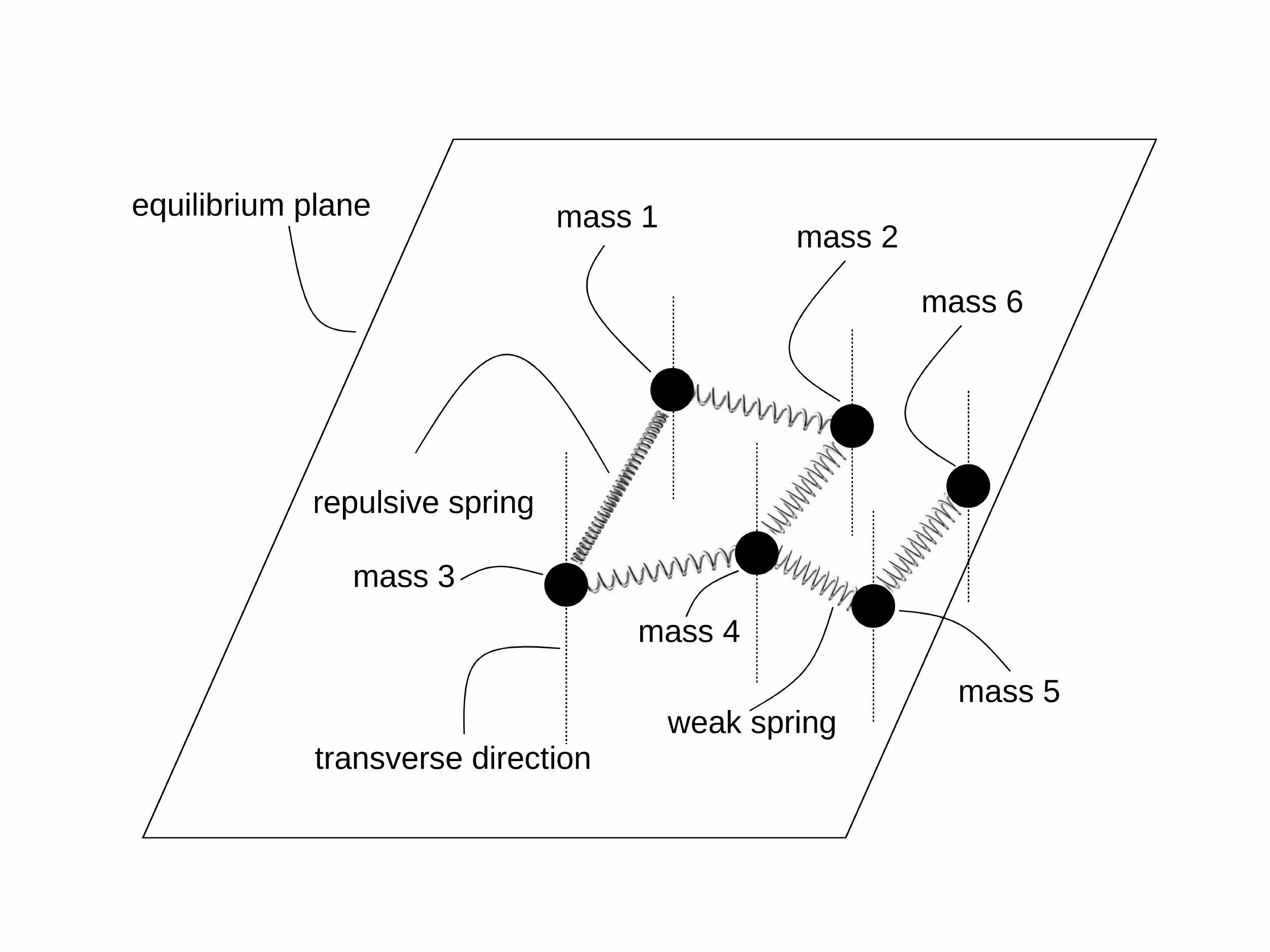}
\caption{Mass-spring system  with repulsive springs.}
\label{fig:m-s}
\end{figure}

Let us consider the mass-spring system in Figure \ref{fig:m-s}, assuming all springs of the same strength, except for the weak spring connecting masses $4$ and $5$, and where one of the springs repulses masses $1$ and $3$. Intuition suggests two alternative partitionings: (a) cutting the weak spring, thus separating the ```tail'' consisting of masses $5$ and $6$, and (b) cutting the repulsive spring and one of the attracting springs, linking mass $3$ or mass $1$ (and $2$) to the rest of the system. Partitioning (a) cuts the weak, but attractive spring; while partitioning (b) cuts one repulsive and one attracting springs of the same absolute strength ``canceling'' each other influence. If the cost function minimized by the partitioning were the total sum of the removed edges, partitioning (a) would be costlier than (b). Within the variants of the partition (b), the most balanced partitioning is the one separating masses $1$ and $2$ from the rest of the system. Let us now examine the Fiedler vectors of the spectral clustering approaches under our consideration. 

The graph corresponding to the mass-spring system in Figure \ref{fig:m-s}, assuming all edges have unit weights, except for the weight $0.2$ of the $(4-5)$ edge, and with $-1$ weight  of the $(1-3)$ edge, 
has the adjacency matrix %and Laplacian matrices as follows,
\begin{equation}\label{e3aa}
 A = \left( \begin{array}{cccccc}
0 & 1 & -1& 0 & 0 & 0\\
1 & 0 & 0 & 1 & 0 & 0\\
-1 & 0 & 0 & 1 & 0 & 0\\
0 & 1 & 1 & 0 & .2 & 0\\
0 & 0 & 0 & .2 & 0 & 1\\
0 & 0 & 0 & 0 & 1 & 0\end{array} \right).
\end{equation}
% { and } 
% \begin{equation}\label{e3a}
% L = \left( \begin{array}{cccccc}
% 0 & -1 & 1& 0 & 0 & 0\\
%   -1 & 2 & 0 & -1 & 0 & 0\\
% 1 & 0 & 0 & -1 & 0 & 0\\
% 0 & -1 & -1 & 2.2 & -.2 & 0\\
% 0 & 0 & 0 & -.2 & 1.2  & -1\\
% 0 & 0 & 0 & 0 & -1 & 1\end{array} \right).
% \end{equation}
% The signed Laplacian, corresponding to the adjacency matrix $A$, is
% \begin{equation}\label{e3b}
% \bar{L} =  \left( \begin{array}{cccccc}
% 2 & -1 & 1& 0 & 0 & 0\\
%   -1 & 2 & 0 & -1 & 0 & 0\\
% 1 & 0 & 0 & -1 & 0 & 0\\
% 0 & -1 & -1 & 2.2 & -.2 & 0\\
% 0 & 0 & 0 & -.2 & 1.2  & -1\\
% 0 & 0 & 0 & 0 & -1 & 1\end{array} \right). 
% \end{equation}

Let us also consider a graph like the one corresponding to the 
 mass-spring system in Figure~\ref{fig:m-s}, but with the repulsive spring eliminated. We nullify the negative weight in the graph adjacency matrix by $A_0=\max(A,0)$ 
 and denote the corresponding to $A_0$ graph Laplacian matrix by $L_0$.
% \begin{equation}\label{e4}
% L_0 =  \left( \begin{array}{cccccc}
% 1 & -1 & 0& 0 & 0 & 0\\
%   -1 & 2 & 0 & -1 & 0 & 0\\
% 0 & 0 & 1 & -1 & 0 & 0\\
% 0 & -1 & -1 & 2.2 & -.2 & 0\\
% 0 & 0 & 0 & -.2 & 1.2  & -1\\
% 0 & 0 & 0 & 0 & -1 & 1\end{array} \right). 
%\end{equation}

\begin{figure}
\centering
\includegraphics[width=0.99\linewidth]{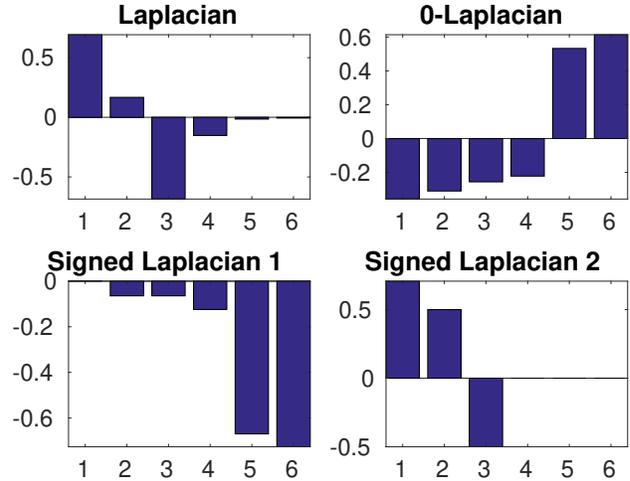}
\caption{Laplacian eigenvectors: the original $L$ (top left), the original with negative weights nullified $L_0$ (top right),
and the signed Laplacian $\bar{L}$ first (bottom right) and second (bottom left) eigenvectors, for a $6$-mass string with a negative weight at one edge between vertices $1$ and $3$.}
\label{fig:negative6masses}
\end{figure}

Figure~\ref{fig:negative6masses} displays the corresponding Fiedler vectors of original $L$ (top left), original with negative weights nullified $L_0$ (top right), and both main modes of the signed Laplacian $\bar{L}$ (bottom). The original Laplacian (top left) suggests meaningful clustering of vertices $1$ and $2$ vs. $3$ and $4$. Dropping the negative weight results in cutting the weakly connected tail of the cobra, see Figure~\ref{fig:negative6masses} top right. The first eigenvector of the signed Laplacian in Figure~\ref{fig:negative6masses} bottom right appears meaningless for clustering, even though it is far from looking as a constant. The second eigenvector of the signed Laplacian in Figure~\ref{fig:negative6masses} bottom left suggests cutting off vertex $3$ from $1$ and $2$, which is not well balanced.

\subsection{``Dumbbell'' graph}
\begin{figure}
\centering
\begin{tikzpicture}[shorten >=1pt,-]
  \tikzstyle{vertex}=[circle,fill=black!25,minimum size=17pt,inner sep=0pt]

  \foreach \name/\angle/\text in {P-1/210/5, P-2/150/6, 
                                  P-3/90/1, P-4/30/2, P-5/-30/3, P-6/-90/4}
    \node[vertex,xshift=6cm,yshift=.5cm] (\name) at (\angle:1cm) {$\text$};

  \foreach \name/\angle/\text in {Q-1/156/8, Q-2/104/7, 
                                  Q-3/52/13, Q-4/0/12, Q-5/-52/11, Q-6/-104/10, Q-7/-156/9}
    \node[vertex,xshift=9cm,yshift=.5cm] (\name) at (\angle:1cm) {$\text$};

  \foreach \from/\to in {1/2,2/3,3/4,4/5,5/1,1/3,2/4,3/5,4/1,5/2,1/6,2/6,3/6,4/6,5/6}
    { \draw (P-\from) -- (P-\to); } 
    \foreach \from/\to in {1/2,2/3,3/4,4/5,5/1,1/3,2/4,3/5,4/1,5/2,1/6,2/6,3/6,4/6,5/6,1/7,2/7,3/7,4/7,5/7,6/7}  
    {\draw (Q-\from) -- (Q-\to); }

  \draw (P-5)--(Q-7);
  \draw (P-6) -- (Q-6);
  \draw [color=red, line width=1mm](P-3) -- (Q-2);
  \draw [color=red, line width=1mm] (P-4) -- (Q-1);
\end{tikzpicture}
\caption{Dumbbell graph, with two negative edges, $(1,7)$ and $(2,8)$, marked thick red.}
\label{fig:dumbbellgraph}
\end{figure}
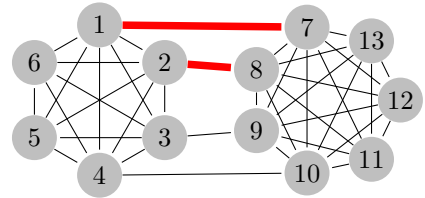
\begin{figure}
\centering
\includegraphics[width=\linewidth]{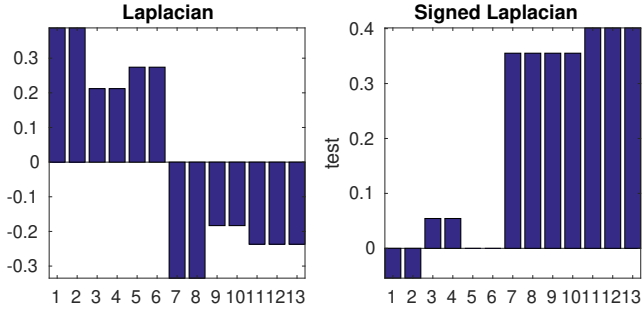}
\caption{Dumbbell graph, eigenvectors of the original Laplacian $L$ (left)
and the signed Laplacian $\bar{L}$ (right)}
\label{fig:dumbbell}
\end{figure}
Our final example is the ``Dumbbell'' graph, displayed in Figure \ref{fig:dumbbellgraph},
 consisting of two complete sub-graphs of slightly unequal sizes, $6$ and $7$, to break the symmetry, attracted by two edges with positive weights, $(3,9)$ and $(4,10)$, and at the same time repelled by two other edges, $(1,7)$ and $(2,8)$, with negative weights, where all weights are unit by absolute value.  Since the weights of the $4$ edges between the two complete sub-graphs 
average to zero, intuition suggests cutting all these $4$ edges, separating the two complete sub-graphs. 

Figure~\ref{fig:dumbbell} displays the corresponding eigenvectors of the original $L$ (left) and the signed Laplacian $\bar{L}$ (right). The signs of the components of the Fiedler vector in the left panel
clearly point to the intuitively expected bisection, keeping the two complete sub-graphs intact. The~eigenvector of the signed Laplacian $\bar{L}$ in Figure~\ref{fig:dumbbell} (right) is quite different and suggest clustering vertices $1$ and $2$, cutting off not only the edges $(1,7)$ and $(2,8)$ with negative weights, but also a large number of edges with positive weights connecting vertices $1$ and $2$ within the first complete sub-graph. The positive components $3$ and $4$ suggest counter-intuitive
cutting off vertices $3$ and $4$ from the first complete sub-graph vertex set $1,\ldots,6$ and cluster them with the vertices $7,\ldots,13$ of the second complete sub-graph, due to the presence of two edges with positive weights, $(3,9)$ and $(4,10)$.

\subsection{Spectral clustering via relaxation}\label{ss:relax}
A common approach to formulate spectral graph partitioning is via relaxation of combinatorial minimization problems, even though it is difficult to mathematically analyze how different cost functions in the combinatorial formulation affect clustering determined via their relaxed versions.  

Let us compare the standard ``ratio cut,'' e.g.,\ \cite{Meila01learningsegmentation,ng2002spectral}, leading to the 
traditional graph Laplacian, and ``signed ratio cut'' of \cite{doi:10.1137/1.9781611972801.49}, used to justify the definition of the signed Laplacian. Let a graph with the set of vertices $V$ be cut into two sub-graphs induced by $X$ and $V\setminus X$. The cut value $Cut(X,V\setminus X)$ is defined as the number of cut edges for unweighted graphs and the sum of the weights of cut edges for weighted graphs. 
In signed graphs, thus, $Cut(X,V\setminus X) = Cut^+(X,V\setminus X) - Cut^-(X,V\setminus X)$, where $Cut^+(X,V\setminus X)$ ($Cut^-(X,V\setminus X)$) denotes the sum of the absolute values of the weights of positive (negative) cut edges. 
The combinatorial balanced graph partitioning is minimizing the ratio of $Cut(X,V\setminus X)$ and the sizes of the partitions; 
its relaxation gives the spectral partitioning using the Fiedler vector of the graph~Laplacian.

%For uniform partitioning, one wants to balance the resulting sub-graphs, which is typically done by dividing the cut value by the sizes of the clusters, giving $RatioCut(X,V\setminus X)=Cut(X,V\setminus X)(|X|^{-1}+|V\setminus X|^{-1})$. %
%\footnote{Substituting sub-graph volumes for sizes leads to the normalized Laplacian, not covered here.}

The signed ratio cut of \cite{doi:10.1137/1.9781611972801.49} is defined by substituting the ``signed cut'' $SignedCut(X,V\setminus X)$ defined as 
$2 Cut^+(X,V\setminus X) + Cut^-(X,X)+ Cut^-(V\setminus X,V\setminus X)$
for the ``cut''. However, the value of all negative edges
$Cut^-(X,V\setminus X) + Cut^-(X,X) + Cut^-(V\setminus X,V\setminus X)$
 in the signed graph remains constant, no matter what $X$ is. We~notice that, up to this constant value,
$SignedCut(X,V\setminus X)$ is equal to 
\[2 Cut^+(X,V\setminus X) - Cut^-(X,V\setminus X).\]
This expression is similar to that of $Cut(X,V\setminus X)$, but the term $Cut^+(X,V\setminus X)$ appears with the multiplier $2$, which suggests that the cuts minimizing quantities involving $SignedCut(X,V\setminus X)$ could tend to ignore the edges with negative weights, focusing instead on cutting the edges with small positive weights. In deep contrast, the positive and negative weights play equal roles in the definition of $Cut(X,V\setminus X)$. 
%It is unclear why minimizing the signed ratio cut would make more sense compared to the standard ratio cut for signed graphs. Moreover, the definition of the signed Laplacian requires exactly \emph{doubling} the term $Cut^+(X,V\setminus X)$, which appears somewhat arbitrary. 

\subsection{Comparing the eigenvectors}
It is challenging to directly quantitatively compare various spectral clustering formulations where the clusters are determined from eigenvectors, since the eigenvectors depend  on matrix coefficients in a complex way. 
We have to rely on simple examples, where we can visualize shapes of the eigenvectors and informally argue which kinds of shapes are beneficial for clustering.  

To add to the trouble, there is apparently still no algorithm universally accepted by experts for an ultimate determination of multiway clusters from several eigenvectors. With this in mind, we restrict ourselves to determining the clusters from the component signs of only one eigenvector---the Fiedler vector for the traditional Laplacian, assuming the corresponding eigenvalues are simple. For the signed Laplacian, the analog of the Fiedler vector is defined in \cite{doi:10.1137/1.9781611972801.49} as corresponding to the smallest, or second smallest, eigenvalue of the signed Laplacian, depending on if the trivial constant eigenvector  is absent.  

In~practice, however, this single eigenvector that determines clustering is computed only approximately, typically being mostly contaminated by other eigenvectors, corresponding to the nearby eigenvalues, especially clustered, so one needs to take into account these other eigenvectors. Our first goal is to check the shapes of several exact eigenmodes already displayed in Figures \ref{fig:2} and \ref{fig:3} and to argue which shapes can be more suitable for automatic partitioning. 

Figure \ref{fig:3} right panel displays the eigenmodes of the signed Laplacian for the 
same weights as in the left panel for the original Laplacian. We observe that, indeed, as we have proved above, 
one of the eigenvectors is piece-wise constant, as in Figure \ref{fig:2} right panel. 

Moreover, the shapes of the other eigenmodes of the signed Laplacian in Figure~\ref{fig:3} right panel also look more 
similar to those in Figure \ref{fig:2} right panel, corresponding to zero weight, than Figure \ref{fig:3} left panel, corresponding to the original graph Laplacian with the same weights. 

The displayed eigenvectors of both the original and signed Laplacian exhibit jumps in the same location of the negative weight in Figure~\ref{fig:3}. However, the jumps are more pronounced in Figure \ref{fig:3} left panel (original Laplacian) due to sharp edges, compared to those in Figure \ref{fig:3} right panel (signed Laplacian), making the location of the former jumps potentially easier to detect automatically than the latter ones, if the eigenvectors are perturbed due to, e.g.,\ numerical inaccuracies.

\begin{figure}
\centering
\includegraphics[width=0.49\linewidth]{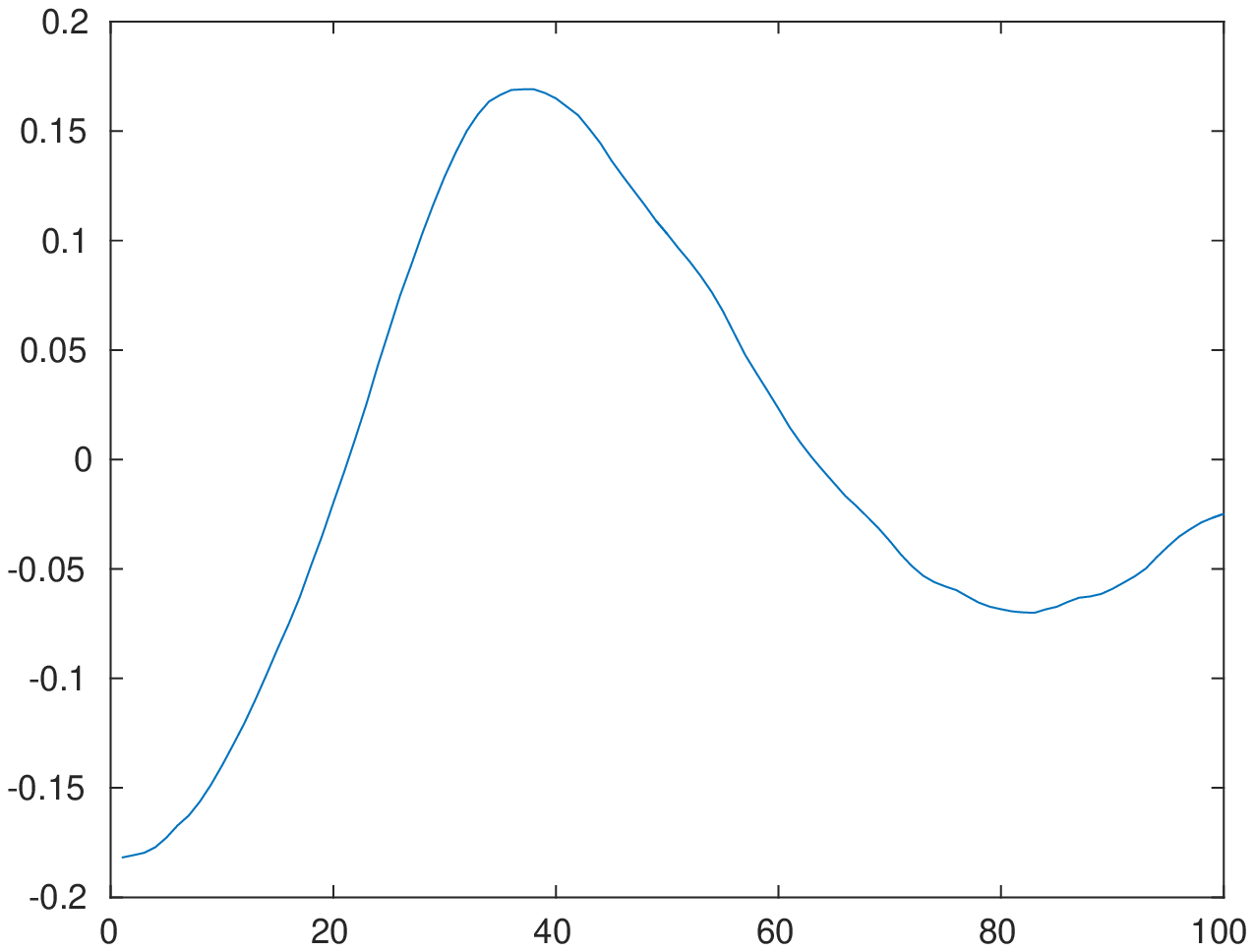}
\includegraphics[width=0.49\linewidth]{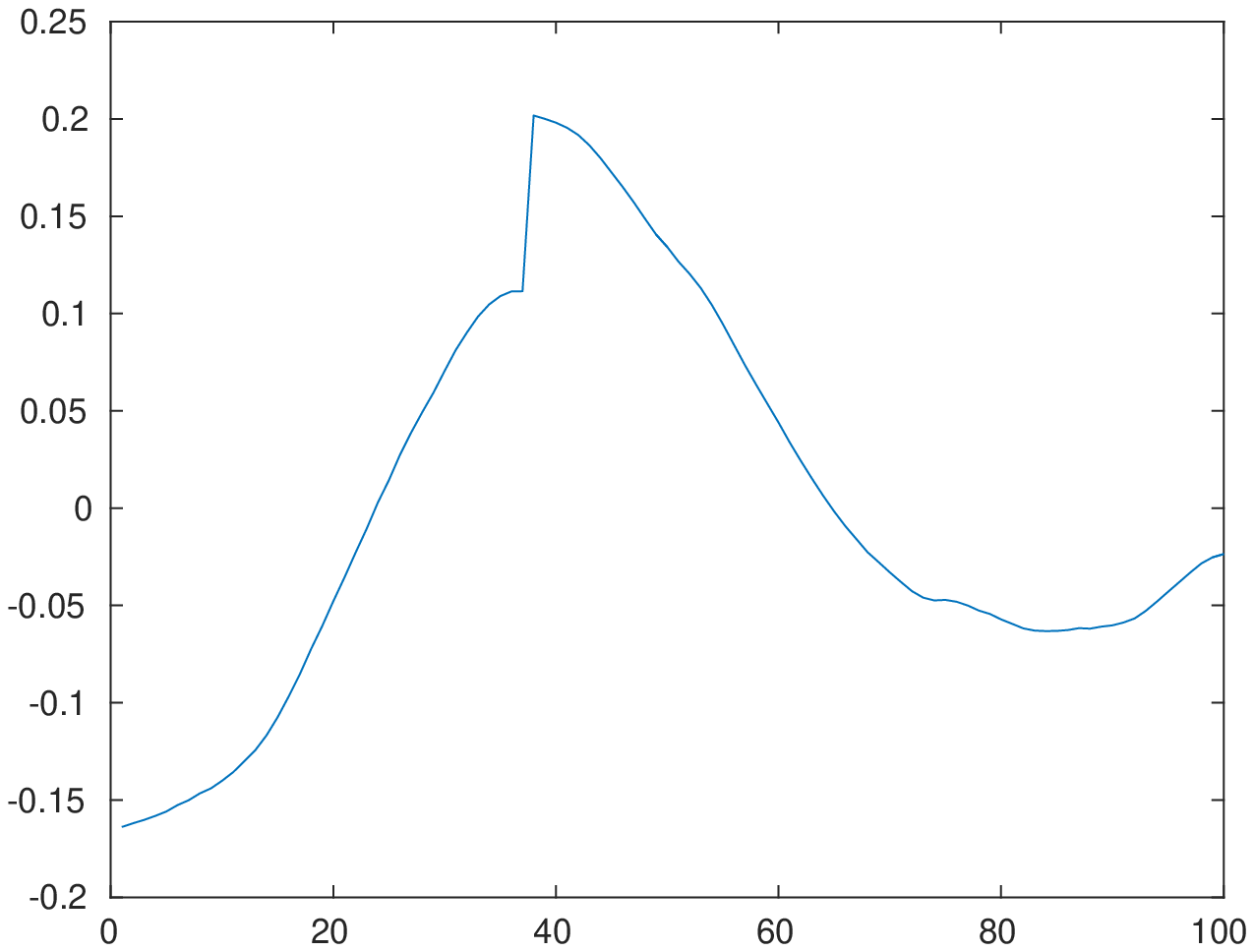}\\
\includegraphics[width=0.49\linewidth]{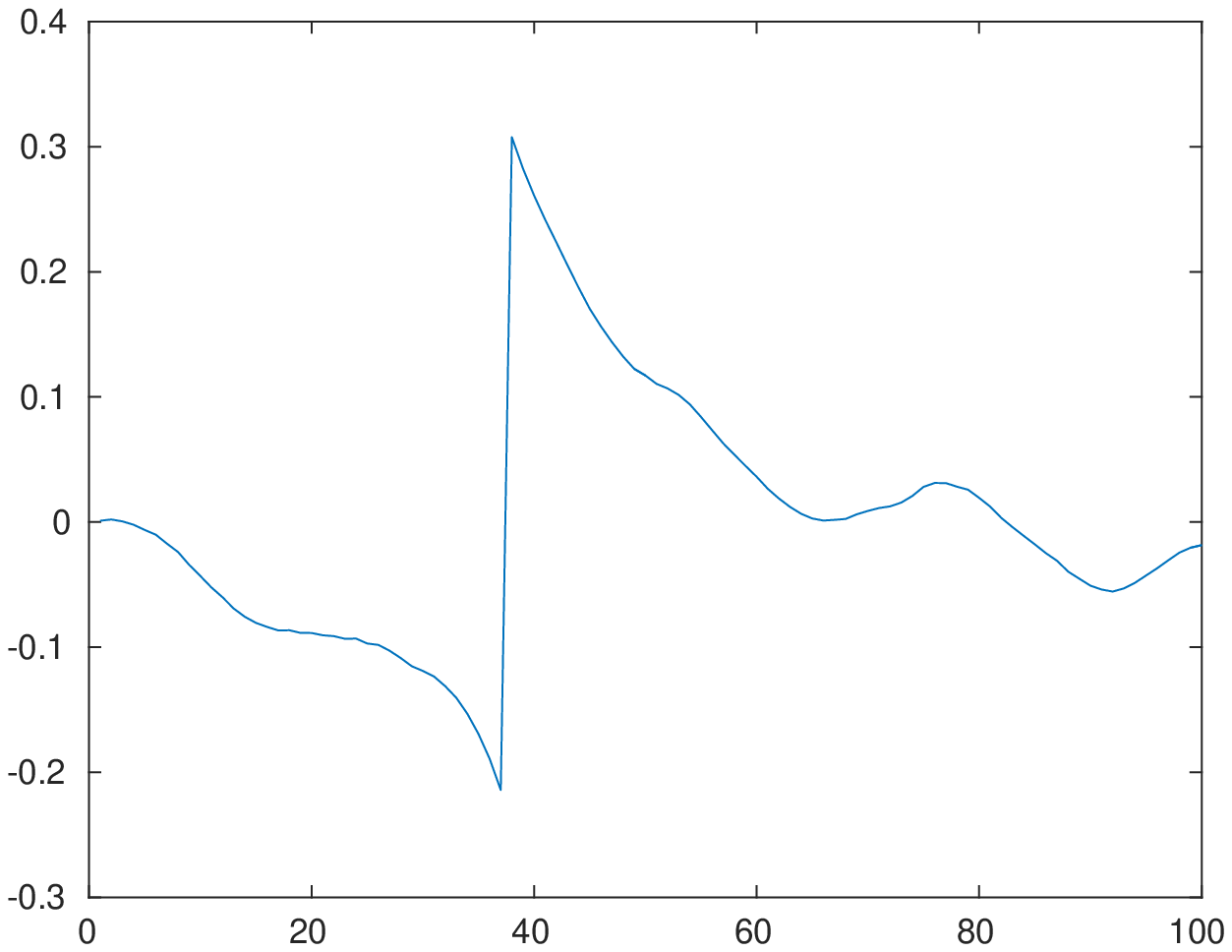}
\includegraphics[width=0.49\linewidth]{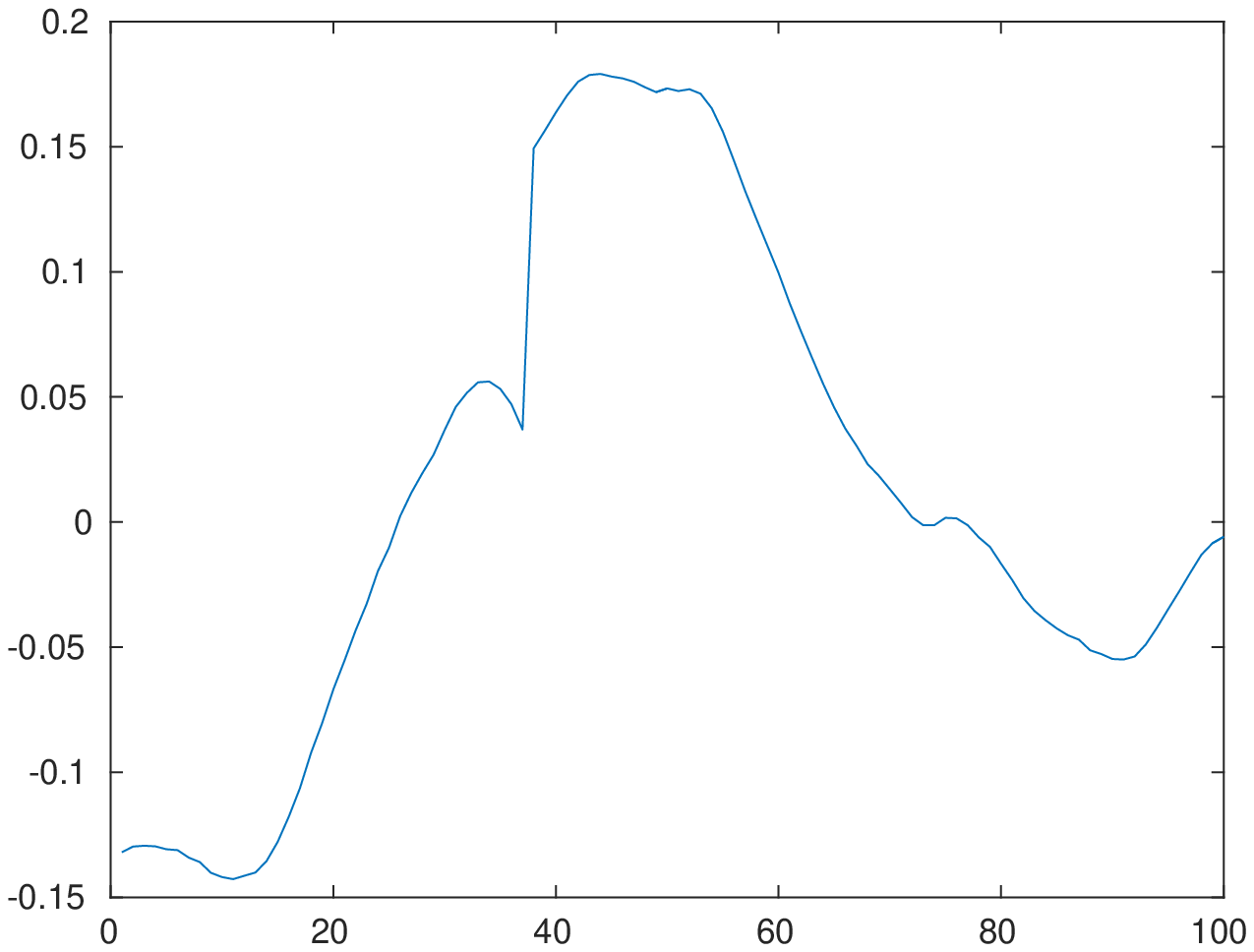}
\caption{Approximate Laplacian eigenmode, unit (a: top left), zero (b: top right), and negative weight at one edge for the original (c: bottom left) and signed (d: bottom right) Laplacians.}
\label{fig:naZero}
\end{figure}

Now we turn our attention to the single eigenvector, but approximated using an iterative eigenvalue/eigenvector solver (eigensolver);
e.g., \cite{k2003,zk17}.
 To set up a direct numerical comparison for our string example, we need to choose a practical eigensolver, so let us briefly discuss computational aspects of spectral clustering. 
 The Fiedler vector, or a group of the eigenvectors, corresponding to the left-most eigenvalues of a symmetric eigenvalue problem needs to be computed iteratively. The~size of the Laplacian matrix is equal to the number of data points, which in modern applications is often extremely large. Most textbook eigensolvers, especially based on matrix transformations, become impractical for large scale problems, where in some cases the Laplacian matrix itself cannot be easily stored, even if it is sparse.   
We follow \cite{k2003} advocating the Locally Optimal Block Preconditioned Conjugate Gradient (LOBPCG) method; see \cite{K01}. LOBPCG does not need to store the matrix $L$ in memory, but requires only the result of multiplying the matrix $L$ by a given vector, or a block of vectors. This characteristic makes LOBPCG applicable to eigenvalue analysis problems of very high dimensions, and results in good parallel scalability to large matrix sizes processed on many parallel processors; e.g.,\ see reference \cite{BLOPEX}, describing our open source and publicly available implementation of LOBPCG. We refer to \cite{zk17} for performance and timing.
 
 Available convergence theory of LOBPCG in \cite{K01} requires the matrix be symmetric, but not necessarily 
 with all non-negative eigenvalues, i.e.,\ a possible presence of negative eigenvalues  still satisfies the convergence assumptions. 
 The calculation of the product of the matrix $L$ by a vector is the main cost per iteration, no matter if the weights are positive or negative. 
 
 We perform $30$ iterations of LOBPCG, without preconditioning and starting from a random initial approximation---the same for various choices of the weights and for different Laplacians for our discrete string example.
The number of iterations is chosen small enough to amplify the influence of 
inaccuracy in approximating the eigenvector iteratively. We display a representative case in Figure \ref{fig:naZero} showing the approximately computed Laplacian eigenmodes with the unit (a), zero (b), and negative (c) weight at one edge, as well as the signed Laplacian~(d), corresponding to the exact eigenfunctions in Figures~\ref{fig:2}~and~\ref{fig:3}. Initial large contributions from other eigenmodes, shown in Figures~\ref{fig:2}~and~\ref{fig:3}, remain unresolved, as anticipated. Two-way partitioning according to the signs of the components of the computed eigenmode of the Laplacian with the negative weight nullified, Figure \ref{fig:naZero} (b), or 
the signed Laplacian, Figure~\ref{fig:naZero}~(d), would result in wrong clusters.

In a sharp contrast, the exact eigenmode (the blue line in Figure~\ref{fig:3} left panel) of the original Laplacian with the negative weight $-0.05$ demonstrates a sharp edge with a large jump between its components of the opposite signs at the correct location of the negative edge, between the $37$ and $38$ vertices. This large jump is inherited by the corresponding approximate eigenmode in Figure \ref{fig:naZero} (c), differentiating it from all other approximate eigenmodes in Figure \ref{fig:naZero}.  The opposite signs of the components of the eigenmode in Figure \ref{fig:naZero} (c) allow determining the correct bisection. Large amplitudes of the absolute values of the components around the jump location in Figure \ref{fig:naZero} (c) make such a determination robust with respect to perturbations and data noise.

There are two reasons why the computed eigenmode in Figure \ref{fig:naZero} (c) visually much better approximates
the exact Fiedler vector compared to other cases in Figure~\ref{fig:naZero}. The first one is that the shape 
 of the exact Fiedler eigenmode (the blue line in Figure~\ref{fig:3} left panel) is pronounced and quite different from those of other eigenfunctions in Figure~\ref{fig:3} left panel. 
The second reason is related to \emph{condition numbers} of eigenvectors, primarily determined by gaps in the matrix spectrum.
 
The convergence speed of iterative approximation to an eigenvector, as well as eigenvector sensitivity with respect to perturbations in the matrix entries, e.g., due to noise in the data, is mostly
determined by a quantity, called the \emph{condition number} of the eigenvector, defined for symmetric matrices as the ratio of the spread of the matrix spectrum to the gap in the eigenvalues. The larger the condition number is, the slower the typical convergence is and more sensitive to the perturbations the eigenvector becomes. 
The trivial zero eigenvalue of the original Laplacian can be excluded from the spectrum, if the influence the corresponding trivial eigenvector, made of ones, may be ignored. 
For the eigenvector corresponding to the smallest nontrivial eigenvalue, the gap is simply the difference between this eigenvalue and the nearest eigenvalue.

What happens in our example, as we see numerically, is that the largest eigenvalue remains basically the same for all variants, so we only need to check the gap. 
It turns out that the gap for the signed Laplacian is about $3$ times smaller, for all tested values of the negative weight, compared to the gap for the case of the zero weight, explaining why we see no improvement in Figures \ref{fig:naZero} (b) and (d), compared to (a). In contrast, introducing the negative weight in the original Laplacian tends to make the target smallest eigenvalue smaller, even negative, in our test for the discrete string, while barely changing the other eigenvalues nearby. As a result, the gap with the negative weight $-0.05$ is $4$ times larger compared to the baseline case of the zero weight. 
We~conclude that the eigenvector condition number for the signed Laplacian is about $3$ times larger, while for the original Laplacian is $4$ times smaller, depending on the negative weight $-0.05$, compared to the baseline  eigenvector condition number for the Laplacian with zero weight. We conclude that in this example the signed Laplacian gives $12$ times larger condition number of the eigenvector of interest and thus is numerically inferior for spectral clustering compared to the original Laplacian. 

\section{Possible extensions for future work}\label{s:future}
We concentrate on the model of the system of masses connected with springs only 
because it directly leads to the standard definition of the graph Laplacian, giving us 
a simple way to justify our introduction of negative weights. 
Similarly, we restrict the vibrations to be transversal, since then we can use the 
classical two-way partitioning definition based on the signs of the components of the Fiedler vector. 
The negative weights can as well be introduced in other models for spectral clustering---we describe two examples below; cf. \cite{AKnegativePatent}.

The first model is based on vibration modes of a wave equation of a system of interacting quasi-particles subjected to vibrations. Each quasi-particle of the vibration model corresponds to one of the data points. Interaction coefficients of the vibration model are determined by pair-wise comparison of the data points. The interaction is attractive/absent/repulsive and the interaction coefficient is positive/zero/negative if the data points in the pair are similar/not comparable/disparate, respectively. The strength of the interaction and the amplitude of the corresponding interaction coefficient represent the level of similarity or disparity.  

The eigenmodes are defined as eigenvectors of an eigenvalue problem resulting from the usual separation of the time and spatial variables. In low-frequency or unstable vibration modes, the quasi-particles are expected to move synchronically in the same direction if they are tightly connected by the attractive interactions, but in the opposite directions if
the interactions are repulsive, or in the complementary directions (where available) if the interaction is absent. 

Compared to the transversal vibrations already considered, where the masses can only move up or down, on the one hand determining the clusters by analyzing the shapes of the vibrations is less straightforward than simply using the signs of the components, but, on the other hand may allow reliable detection of more than two clusters from a single eigenmode. 
For example, a quasi-particle representing an elementary volume of an elastic body in three-dimensional space has six degrees of freedom, which may allow definition of up to twelve clusters from a single vibration mode. 
Multiway algorithms %, such as in \cite{310898,doi:10.1093/comnet/cnt021,AAAI1612307}, 
of spectral graph partitioning have to be adapted to this case, where a quasi-particle associated with a graph vertex has multiple degrees of freedom. 

A second, alternative, model is a system of interacting quasi-particles subjected to concentration or diffusion, described by  concentration-diffusion equations. Every quasi-particle of the concentration-diffusion model corresponds to a point in the data. Conductivity coefficients of interactions of the quasi-particles are determined by pair-wise comparison of data points. The interaction is diffusive and the interaction conductivity coefficient is positive if the data points in the pair are similar. The interaction is absent and the interaction conductivity coefficient is zero if the data points in the pair are not comparable. Finally, the interaction is concentrative and the interaction conductivity coefficient is negative if the data points in the pair are disparate. The strength of the interaction and the amplitude of the interaction coefficient represent the level of similarity or disparity. 

As in the first model, the eigenvalue problem is obtained by the separation of the time and spatial variables in the time dependent diffusion equation. The clusters are defined by the quasi-particles that concentrate together in unstable or slowest eigenmodes,
corresponding to the left part of the spectrum. 

A forward-and-backward diffusion in \cite{1021076,Tang2016} provides a different interpretation of a similar diffusion equation, but the negative sign in the conductivity coefficient is moved to the time derivative, reversing the time direction. Here, the time is going forward (backward) on the graph edges with the positive (negative) weights. Having the time forward and backward in different parts of the same model seems unnatural. 

Finally, our approach allows reversing the signs of all weights, thus treating the minimum cut and the maximum cut problems in the same manner, e.g., applying the same spectral clustering techniques to the original Laplacian, in contrast to the signed Laplacian.   

\section{Conclusions}
Spectral clustering has been successful in many applications, ranging from traditional resource allocation, image segmentation,  and information retrieval to more recent bio- and material-informatics, providing good results at a reasonable cost. 
Improvements of cluster quality and algorithm performance are important, e.g., for big data or real-time clustering. 
We introduce negative weights in the graph adjacency matrix for incorporating disparities in data via spectral clustering that traditionally only handles data with similarities. 

Incorporating the disparities in the data into spectral clustering is expected to be of significance and 
have impact in any application domain where the data disparities naturally appear, e.g., if the data comparison involves  correlation or covariance. If data features are represented by elements of a vector space equipped with a vector scalar product, the scalar product can be used for determining the pair-wise comparison function having both negative and non-negative values. 

Traditional spectral clustering, with only non-negative weights, remains largely intact when negative weights are introduced. Eigenvectors corresponding to the algebraically smallest eigenvalues (that can be negative) of the graph Laplacian define clusters of higher quality, compared to those obtained via the signed Laplacian.
The mass-spring system with repulsive springs justifies well the use of the standard Laplacian for clustering, in contrast to the signed Laplacian that may result in counter-intuitive partitions.

\bibliographystyle{siamplain}
%%\bibliographystyle{IEEEbib}
%\bibliography{refsnew} 

\end{document}